\documentclass[preprint,showpacs,preprintnumbers,amsmath,amssymb]{revtex4}
\usepackage{graphicx}
\usepackage{dcolumn} 
\usepackage{bm}      
\begin{document}
\title{Single bunch transverse instability
in a circular accelerator \\
with chromaticity and space charge}
\author{V. Balbekov}
\affiliation {Fermi National Accelerator Laboratory\\
P.O. Box 500, Batavia, Illinois 60510}
\email{balbekov@fnal.gov} 
\date{\today}
%%%%%%%%%%%%%%%%%%%%%%%%%%%%%%%%%%%%%%%%%%%%%%%%%%%%%%%%%%%%%%
\begin{abstract}

The transverse instability of a bunch in a circular accelerator  
is elaborated in this paper. 
A new tree-modes model is proposed and developed to describe 
the most unstable modes of the bunch.
This simple and flexible model includes chromaticity and space charge, and
can be used with any bunch and wake forms.
The dispersion equation for the bunch eigentunes is obtained in form 
of a third-order algebraic equation.
The known head-tail and TMCI modes appear as the limiting cases which are 
distinctly bounded at zero chromaticity only.
It is shown that the instability parameters depend only slightly
on the bunch model but they are rather sensitive to the wake shape.
In particular, space charge effects are investigated in the paper 
and it is shown that their influence depends on sign of wake field
enhancing the bunch stability if the wake is negative.
The resistive wall wake is considered in detail including a comparison 
of single and collective effects.
A comparison of the results with earlier publications is carried out.

\end{abstract}
%%%%%%%%%%%%%%%%%%%%%%%%%%%%%%%%%%%%%%%%%%%%%%%%%%%%%%%%%%%%%%
\pacs{29.27.Bd} 
\maketitle

-----------------------------------------------------------------
-------------
\\
\begin{center}
CONTENTS                                     
\end{center}
\noindent
I.~~~~~Introduction
\\
II.~~~~Basic equations and definitions
\\
III.~~~Exact solution for hollow bunch
\\
IV.~~~Integral equations for multipoles
\\
V.~~~~Tree-modes model with chromaticity
\\
VI.~~~Realistic bunch with arbitrary wake
\\
VIA.~Short rectangular wake
\\
VIB.~Resistive wake
\\
VII.~~Space charge effects
\\
VIIA.~Discussion
\\
VIII.~Conclusion
\\
References
\\
Appendix
\\
------------------------------------------------------------------------------

%
%%%%%%%%%%%%%%%%%%%%%%%%%%%%%%%%%%%%%%%%%%%%%%%%%%%%%%%%%%%%%%   SEC.1
\section {Introduction}
%%%%%%%%%%%%%%%%%%%%%%%%%%%%%%%%%%%%%%%%%%%%%%%%%%%%%%%%%%%%%%   SEC.1   
%

Two kinds of single bunch transverse  instability in circular accelerators 
are distinguished at present: the head-tail one and the transverse mode 
coupling instability (TMCI).
First of them has been studied first by Pellegrini \cite{PEL}
and Sands \cite{SAND} who were treating the transverse bunch oscillations as a 
combination  of uncoupled multipoles $\,\propto \exp(im\phi)\,$ with $\,\phi\,$ 
as synchrotron phase. 
Role of synchrotron amplitude has been investigated later by Sacherer who was 
introducing the conception of radial modes \cite{SAH}.
A lot of papers dealing with this were published later,
including reviews and books (see e.g. \cite{CHAO,NG}).
The most important conclusion is that the head-tail instability is possible at 
non-zero chromaticity only and depends on its sign.

Other type of instability was observed in many electron machines regardless of
chromaticity and being primary referred as ``transverse turbulence''.
Its first explanation has been proposed by Kohaupt who used a simple model 
of two particles propelling each other through the mediation of constant 
wake fields \cite{KOH}. 
Later the theory has been evolved on the base of Vlasov equation treating
the instability as a result of reciprocal influence of neighboring pairs 
of the head-tail modes which frequencies approach each other
due to a wake field, whence the term TMCI appears~\cite{ZOT,CHIN,CHAO,NG}.
Chromaticity has been ignored in these investigations as a 
factor of secondary importance. 
Some space charge effects has been considered later 
for both head-tail and TMCI instabilities \cite{BL,BU,B1}. 
However, specific models like the hollow bunch in the square potential well 
have been used in \cite{BL}, and only high space charge limit was 
investigated in Ref.~\cite{BL,BU,B1}.

An important observation is that, in all considered cases, 
the lowest multipoles $m=0,\,\pm1$ or their combinations are the most prone
to instability and pose a major threat to the accelerator operation.
It is shown in this paper that the general theory can be essentially simplified 
and unified being restricted by these cases.

General 3-modes model is developed for this, providing simple analytical 
formulae for all types of single bunch instability.
The model is built on the base of exact solutions for hollow bunch which are
obtained in the paper without any additional assumptions like truncated 
multipole expansion.
The dangerous modes are separated and extended to realistic 
distributions with arbitrary wake, chromaticity and space charge included.

It is shown that the instability parameters depend only slightly on the 
bunch shape being rather sensitive to the wake function form.
Resistive wall wake is investigated especially to assign which effect is more 
dangerous in specific cases: either TMCI or collective mode instability.
Chromaticity is taken into account in all the cases demonstrating an absence
of clearly defined boundary between the head-tail 
instability (separated multipoles) and TMCI (coalesced multipoles).
Space charge effects are studied with arbitrary ratio of betatron tune 
shift to synchrotron tune. 
In contrast with earlier papers \cite{BL,BU,B1}, the investigation is not 
restricted by simplified bunch models or by limiting cases only.
It is shown that the space charge effect can be either advantageous or not 
for the bunch stability, dependent on the wake sign.

%
%%%%%%%%%%%%%%%%%%%%%%%%%%%%%%%%%%%%%%%%%%%%%%%%%%%%%%%%%%%%%%%   SEC.2
\section
{Basic equations and definitions \cite{B1}}
%%%%%%%%%%%%%%%%%%%%%%%%%%%%%%%%%%%%%%%%%%%%%%%%%%%%%%%%%%%%%%%   SEC.2
%

Linear synchrotron oscillations are considered in this paper
being characterized by amplitude $\,A\,$ and phase $\,\phi$, 
or by corresponding Cartesian coordinates: 
%%%%%%%%%%%%%%%%%%%%%%%%%%%%%%%%%%%%%%%%%%%%%%%%%%%%%%%%%%%%%%%   1    
\begin{equation}
 \theta= A\cos\phi, \qquad u=A\sin\phi.
\end{equation}                         
%%%%%%%%%%%%%%%%%%%%%%%%%%%%%%%%%%%%%%%%%%%%%%%%%%%%%%%%%%%%%%%   1
Thus $\,\theta\,$ is azimuthal deviation of the particle from the bunch center
in the rest frame whereas the variable $u$ is proportional to the momentum 
deviation with respect to the central bunch momentum 
(proportionality coefficient is not a factor in this paper). 
Steady state of a bunch will be described by the distribution function 
$\,F(A)\,$ and by corresponding linear density
%%%%%%%%%%%%%%%%%%%%%%%%%%%%%%%%%%%%%%%%%%%%%%%%%%%%%%%%%%%%%%%   2    
\begin{equation}
 \rho(\theta)=\int_{-\infty}^{\infty}F(\sqrt{\theta^2+u^2})\,du
\end{equation}                         
%%%%%%%%%%%%%%%%%%%%%%%%%%%%%%%%%%%%%%%%%%%%%%%%%%%%%%%%%%%%%%%   2
%
with the normalization conditions
%
%%%%%%%%%%%%%%%%%%%%%%%%%%%%%%%%%%%%%%%%%%%%%%%%%%%%%%%%%%%%%%%   3    
\begin{equation}
 2\pi\int_0^\infty F(A)\,AdA =1,\qquad 
 \int_{-\infty}^\infty \rho(\theta)\,d\theta=1.
\end{equation}                         
%%%%%%%%%%%%%%%%%%%%%%%%%%%%%%%%%%%%%%%%%%%%%%%%%%%%%%%%%%%%%%%   3
%
It is convenient to present coherent transverse displacement 
of the bunch in the point of longitudinal phase space $(A,\phi)$
in the form \cite{B1}
%%%%%%%%%%%%%%%%%%%%%%%%%%%%%%%%%%%%%%%%%%%%%%%%%%%%%%%%%%%%%%%   4    
\begin{equation}
 Y(A,\phi)\exp\big[-i(Q_0+\zeta)\,\theta-i\,(Q_0+\nu)\,\Omega_0t\,\big]
\end{equation}                         
%%%%%%%%%%%%%%%%%%%%%%%%%%%%%%%%%%%%%%%%%%%%%%%%%%%%%%%%%%%%%%%   4
where $\,\Omega_0\,$ is the revolution frequency, $\,Q_0$ is the central 
betatron tune,  $\,\nu\,$ is an addition due to a wake field, 
and $\,\zeta\,$~is the normalized chromaticity:
%%%%%%%%%%%%%%%%%%%%%%%%%%%%%%%%%%%%%%%%%%%%%%%%%%%%%%%%%%%%%%%   5
\begin{equation}
 \zeta=\frac{\Omega_0Q'_p}{\Omega'_p}=\frac{\xi}{1/\gamma^2-\alpha},     
\end{equation}
%%%%%%%%%%%%%%%%%%%%%%%%%%%%%%%%%%%%%%%%%%%%%%%%%%%%%%%%%%%%%%%   5
with $\xi$ as usual chromaticity, and $\alpha$ as the momentum 
compaction factor. 
As it has been shown in Ref.~\cite{B1}, the function $\,Y$ satisfies 
the equation
%
%%%%%%%%%%%%%%%%%%%%%%%%%%%%%%%%%%%%%%%%%%%%%%%%%%%%%%%%%%%%%%%   6
\begin{eqnarray}
 \nu Y+i\,Q_s\frac{\partial Y}{\partial\phi}
+\Delta Q_{av} \biggl(\rho(\theta)Y(\theta,u)-\int_{-\infty}^\infty 
 F(\theta,u)Y(\theta,u)du\biggr)   \nonumber\\
=2\int_\theta^\infty
 q(\theta'-\theta)\exp\big[i(\zeta\!-\!\nu)(\theta\!-\!\theta')\big]d\theta'  
 \int_{-\infty}^\infty F(A')Y(A',\phi')\,du'            
\end{eqnarray}
%%%%%%%%%%%%%%%%%%%%%%%%%%%%%%%%%%%%%%%%%%%%%%%%%%%%%%%%%%%%%%%   6
%
where $\,Q_s\,$ is synchrotron tune, 
$\Delta Q_{av}$ is betatron space charge tune shift 
averaged on all variables, and $\,q(\theta)\,$ is the reduced wake 
function which is connected with usual wake field function $W_1(z)$ 
by the relation:  
%%%%%%%%%%%%%%%%%%%%%%%%%%%%%%%%%%%%%%%%%%%%%%%%%%%%%%%%%%%%%%%   7
\begin{equation}
 q(\theta) =\frac{r_0R N W_1(-R\theta)}{8\pi\beta\gamma Q_0} 
\end{equation}
%%%%%%%%%%%%%%%%%%%%%%%%%%%%%%%%%%%%%%%%%%%%%%%%%%%%%%%%%%%%%%%   7
with $\,r_0=e^2/mc^2\,$ as classic radius of the particle, $R$ as the 
accelerator radius, $N$ as the bunch population, $\,\beta\,$ and $\,\gamma\,$ 
as normalized velocity and energy.
Definition of the wake function and numerous examples can be found  
in Ref.~\cite{HAND}. 
More often than not, this function is negative 
(resistive wall wake can be mentioned as the best known example).
Just such a case was considered in the most of published papers.
However, positive wakes are known as well, for example the field created 
by heavy positive ions in a proton beam \cite{HER}.
It will be shown in Sec.VII that the wake sign is especially important when 
space charge is included in the consideration. 
Alternating wakes are possible as well (resonator models \cite{HAND})
but this case is beyond the scope of the paper. 

It is assumed in any case that $\,q(\theta)\,$ is rather short range function 
so that the wake field cannot reach subsequent bunches or turns.
Constant wake $\,q=q_0\,$ will be used to start with,
and more general cases will be considered in Sec.~6.
Besides of these, space charge is not taken into account in 3 nearest 
sections.
With this reservations, Eq.~(6) obtains the simpler form
%%%%%%%%%%%%%%%%%%%%%%%%%%%%%%%%%%%%%%%%%%%%%%%%%%%%%%%%%%%%%%%   8
\begin{eqnarray}
 \nu Y+i\,Q_s\frac{\partial Y}{\partial\phi}=2q_0
 \exp(i\zeta_\nu\theta)                        
 \int_\theta^\infty\!\!\exp(-i\zeta_\nu\theta')
 d\theta' \int_{-\infty}^\infty F(A')Y(A',\phi')\,du'             
\end{eqnarray}
%%%%%%%%%%%%%%%%%%%%%%%%%%%%%%%%%%%%%%%%%%%%%%%%%%%%%%%%%%%%%%%   8
where $\zeta_\nu=\zeta-\nu$. 
Note that, in most cases, the addition $\,\nu\,$ is rather small  
to identify $\zeta_\nu$ with normalized chromaticity given by Eq.~(5).

%%%%%%%%%%%%%%%%%%%%%%%%%%%%%%%%%%%%%%%%%%%%%%%%%%%%%%%%%%%%%%%   SEC.3
%
\section{Exact solutions for hollow bunches}
%
%%%%%%%%%%%%%%%%%%%%%%%%%%%%%%%%%%%%%%%%%%%%%%%%%%%%%%%%%%%%%%%   SEC.3

The hollow bunch model is characterized by the distribution functions
%
%%%%%%%%%%%%%%%%%%%%%%%%%%%%%%%%%%%%%%%%%%%%%%%%%%%%%%%%%%%%%%%   8
\begin{equation}
 F(A)=\frac{\delta(A-A_0)}{2\pi A_0},\qquad  
\rho(\theta)=\frac{1}{\pi\sqrt{\theta_0^2-\theta^2}}
\end{equation}
%%%%%%%%%%%%%%%%%%%%%%%%%%%%%%%%%%%%%%%%%%%%%%%%%%%%%%%%%%%%%%%   8
%
where $\,A_0=\theta_0\,$ is synchrotron amplitude of any particle
and, simultaneously, the bunch half-length.
The model was repeatedly investigated among others   
in frames of separated or truncated multipole approximations
(see e.g. \cite{D2}).  
However, its special interest is just that the exact solution can be 
obtained without any similar assumptions.
Therefore corresponding results will be substantially used in this paper to
develop adequate approximate methods with more realistic distributions,
and to control their accuracy. 

The only amplitude $A=A_0$ is essential in this case. 
Therefore Eq.~(8) with $\zeta_\nu=0$ can be reduced to the one-dimension 
equation for new function $Y(\phi)\equiv Y(A_0,\phi)$: 
%%%%%%%%%%%%%%%%%%%%%%%%%%%%%%%%%%%%%%%%%%%%%%%%%%%%%%%%%%%%%%%   10
\begin{equation}
 \nu Y(\phi)+i\,Q_s Y'(\phi) = \frac{q_0}{\pi}
 \int_{-[\phi]}^{[\phi]} Y(\phi')\,d\phi'   
\end{equation}
%%%%%%%%%%%%%%%%%%%%%%%%%%%%%%%%%%%%%%%%%%%%%%%%%%%%%%%%%%%%%%%   10
where $\,[\phi]\,$~is the periodic polygonal function of period $2\pi$ 
taking the value $\,[\phi]=|\phi|\,$ at $\,|\phi|<\pi$.
It is convenient to separate even and odd parts of the function presenting 
it in the form
%%%%%%%%%%%%%%%%%%%%%%%%%%%%%%%%%%%%%%%%%%%%%%%%%%%%%%%%%%%%%%%   
$
\; Y(\phi) = Y_+(\phi)+Y_-(\phi)\;\;{\rm with}\;\;
 Y_+(\phi)=Y_+(-\phi)\;\;{\rm and}\;\;Y_-(\phi)=-Y_-(-\phi).\;
$
%%%%%%%%%%%%%%%%%%%%%%%%%%%%%%%%%%%%%%%%%%%%%%%%%%%%%%%%%%%%%%%   
%
It is easy to see that the even part satisfies the equation
%
%%%%%%%%%%%%%%%%%%%%%%%%%%%%%%%%%%%%%%%%%%%%%%%%%%%%%%%%%%%%%%%   11
\begin{equation}
 Q_s^2 Y_+''(\phi)+\nu^2Y_+(\phi)
=\frac{2q_0\nu}{\pi}\int_0^{[\phi]} Y_+(\phi')d\phi'  
\end{equation}
%%%%%%%%%%%%%%%%%%%%%%%%%%%%%%%%%%%%%%%%%%%%%%%%%%%%%%%%%%%%%%%   11
%
Restricting the consideration to the region $0<\phi<\pi$, 
one can transform it to ordinary third order differential equation
%
%%%%%%%%%%%%%%%%%%%%%%%%%%%%%%%%%%%%%%%%%%%%%%%%%%%%%%%%%%%%%%%   12
\begin{equation}
 Y_+'''(\phi)+\frac{\nu^2}{Q_s^2}Y_+'(\phi)
=\frac{2q_0\nu}{\pi Q_s^2}Y_+(\phi)
\end{equation}
%%%%%%%%%%%%%%%%%%%%%%%%%%%%%%%%%%%%%%%%%%%%%%%%%%%%%%%%%%%%%%%   12
%
with the boundary conditions
%
%%%%%%%%%%%%%%%%%%%%%%%%%%%%%%%%%%%%%%%%%%%%%%%%%%%%%%%%%%%%%%%   13
\begin{equation}
 Y_+'(0)=0,\qquad Q_s^2Y_+''(0)+\nu^2Y_+(0)=0,\qquad Y_+'(\pi)=0.
\end{equation}
%%%%%%%%%%%%%%%%%%%%%%%%%%%%%%%%%%%%%%%%%%%%%%%%%%%%%%%%%%%%%%%   13
%
Eq.~(12)-(13) have been obtained first in Ref.~\cite{D2} where
graphical solution id displayed as well. 
The analytical solution is represented below in the standard form:
%
%%%%%%%%%%%%%%%%%%%%%%%%%%%%%%%%%%%%%%%%%%%%%%%%%%%%%%%%%%%%%%%   14
\begin{equation}
 Y_+(\phi) = \sum_{j=1}^3 C_j \exp\biggl(\frac{\nu\lambda_j}{Q_s}\biggr)
\end{equation}
%%%%%%%%%%%%%%%%%%%%%%%%%%%%%%%%%%%%%%%%%%%%%%%%%%%%%%%%%%%%%%%   14
%
where $\lambda_{1-3}$ are the roots of the algebraic equation:
%
%%%%%%%%%%%%%%%%%%%%%%%%%%%%%%%%%%%%%%%%%%%%%%%%%%%%%%%%%%%%%%%   15
\begin{equation}
 \lambda^3+\lambda=g,\qquad g=\frac{2q_0 Q_s}{\pi\nu^2}.
\end{equation}
%%%%%%%%%%%%%%%%%%%%%%%%%%%%%%%%%%%%%%%%%%%%%%%%%%%%%%%%%%%%%%%   15
%
The constants $C_{1-3}$ have to be determined 
trough the boundary conditions (13).
The substitution provides a series of linear uniform 
equation for $C_{1-3}$ which is solvable if corresponding determinant 
is equal to zero.
It results in the dispersion equation for the bunch eigentunes~$\nu$:
%%%%%%%%%%%%%%%%%%%%%%%%%%%%%%%%%%%%%%%%%%%%%%%%%%%%%%%%%%%%%%%   16
\begin{eqnarray}
 \lambda_1(\lambda_2-\lambda_3)(1-\lambda_2\lambda_3)
 \exp(\pi\nu\lambda_1/Q_s)\;+                              \nonumber\\
 \lambda_2(\lambda_3-\lambda_1)(1-\lambda_3\lambda_1)
 \exp(\pi\nu\lambda_2/Q_s)\;+                                       \\
 \quad\lambda_3(\lambda_1-\lambda_2)(1-\lambda_1\lambda_2)
 \exp(\pi\nu\lambda_3/Q_s)\;=\,0.\hspace{-4mm}          \nonumber
\end{eqnarray}
%%%%%%%%%%%%%%%%%%%%%%%%%%%%%%%%%%%%%%%%%%%%%%%%%%%%%%%%%%%%%%%   16
%
Thus, the following steps are evident for handling the problem:
(i) to find the roots $\lambda_{1-3}$ of the first Eq.~(15) with arbitrary 
chosen parameter $g$;
(ii) to use them for the solution of Eq.~(16);  
(iii) to substitute the found roots $\,\nu\,$ in second Eq.~(15) obtaining 
dependence $\,\nu(q_0)\,$ by exclusion of the parameter $\,g$.

Generally, this way is applicable for all roots including complex ones. 
However, in this section we will restrict our consideration to real roots 
only, postponing discussing of complex eigentunes to next parts 
where they will be analyzed with chromaticity taken into account.

Eq.~(16) has infinite number of solutions some of which are plotted 
in Fig.~1.
It is seen that $\,\nu\simeq mQ_s\,$ at small $\,q_0\,$ 
which case has to be treated as independent oscillations of different 
multipoles.
Because these tunes are real numbers, instability is impossible at small 
wake and zero chromaticity.
Complex roots appear for the first time at $\,|q_0|>0.567\,Q_s$ as a result 
of the coalescing of multipoles $m=0$ and $m=\pm1$, dependent on sign of 
the wake.
Presented inequality should be treated as the threshold of the lowest 
TMCI mode.
Higher TMCI modes are possible too being caused by a coalescence of higher 
multipoles. 
They have essentially larger thresholds as it is shown in Table I
(only positive wakes are considered in the Table because the picture is 
symmetric). 
No other coalescences and TMCI appearances have been observed at the
calculations.  
%
%%%%%%%%%%%%%%%%%%%%%%%%%%%%%%%%%%%%%%%%%%%%%%%%%%%%%%%%%%%%%%%   Fig.1
\begin{figure}[t!]
\includegraphics[width=85mm]{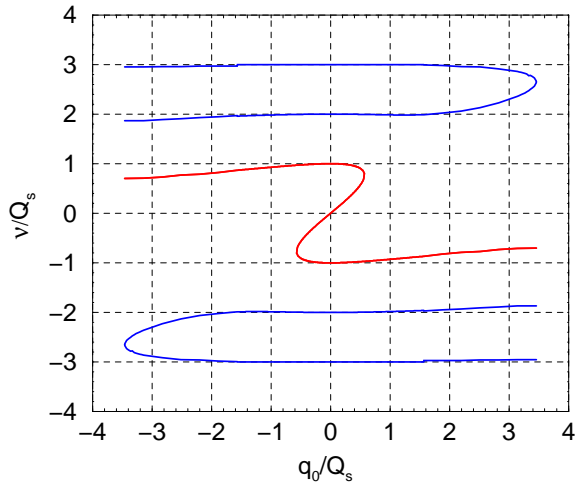}
\caption{Real eigentunes of hollow bunch without chromaticity 
(exact solutions).
 There are complex roots at $|q_0|/Q_s>0.567$ (not shown).}
\end{figure}
%%%%%%%%%%%%%%%%%%%%%%%%%%%%%%%%%%%%%%%%%%%%%%%%%%%%%%%%%%%%%%%   Fig.1
%
%%%%%%%%%%%%%%%%%%%%%%%%%%%%%%%%%%%%%%%%%%%%%%%%%%%%%%%%%%%%%%%   Table 1
\begin{table}[h!]
\begin{center}
\caption{Threshold of different TMCI modes}
\vspace{5mm}
\begin{tabular}{|c|c|c|c|c|c|c|}
\hline 
Coalesced multipoles& (0-1) & (2-3) & (4-5) & (6-7)& (8-9)& (10-11) \\
\hline
$(q_0)_{thresh}/Q_s$&~0.5671~&~3.459~&~7.366~&11.894&16.871& 22.198  \\
\hline
\end{tabular}
\end{center}
\vspace{-3mm}
\end{table}
%%%%%%%%%%%%%%%%%%%%%%%%%%%%%%%%%%%%%%%%%%%%%%%%%%%%%%%%%%%%%%    Table 1
%

Hence the modes $|m|>1$ have a minor interest from point of view of the 
instability threshold.
Therefore our goal should be an investigation of the lower modes presented 
by red line in Fig.~1, using realistic distributions and wake 
functions, with chromaticity and space charge taken into account. 

%
%%%%%%%%%%%%%%%%%%%%%%%%%%%%%%%%%%%%%%%%%%%%%%%%%%%%%%%%%%%%%%%   SEC. 4

\section {Integral equations for multipoles}
 
%%%%%%%%%%%%%%%%%%%%%%%%%%%%%%%%%%%%%%%%%%%%%%%%%%%%%%%%%%%%%%%   SEC. 4
%

General solution of Eq.~(8) can be presented as the Fourier series:
%
%%%%%%%%%%%%%%%%%%%%%%%%%%%%%%%%%%%%%%%%%%%%%%%%%%%%%%%%%%%%%%%   17
\begin{equation}
 Y(A,\phi) = \sum_m Y_m(A)\exp(im\phi)                           
\end{equation}
%%%%%%%%%%%%%%%%%%%%%%%%%%%%%%%%%%%%%%%%%%%%%%%%%%%%%%%%%%%%%%%   17
%
which is just expansion over the multipoles.
Its substitution to Eq.~(8) allows to get set of coupled integral 
equations for the functions $\,Y_m(A)$ \cite{SAH}:
%
%%%%%%%%%%%%%%%%%%%%%%%%%%%%%%%%%%%%%%%%%%%%%%%%%%%%%%%%%%%%%%%   18
\begin{eqnarray}
 (\nu-mQ_s)\,Y_m(A)                               
 =\;2\pi q_0\sum_n\int_0^\infty K_{m,n}(A,A')Y_n(A')\,F(A')A'\,dA' 
\end{eqnarray}
%%%%%%%%%%%%%%%%%%%%%%%%%%%%%%%%%%%%%%%%%%%%%%%%%%%%%%%%%%%%%%%   18
%
with the kernels
%
%%%%%%%%%%%%%%%%%%%%%%%%%%%%%%%%%%%%%%%%%%%%%%%%%%%%%%%%%%%%%%%   19
\begin{eqnarray}
 K_{m,n}(A,A')=\frac{2}{\pi^2}\int_{-A}^A
 \exp(i\zeta_\nu\theta)\,\frac{T_m(\theta/A)
 \,d\theta}{\sqrt{A^2-\theta^2}}                      
 \int_{\theta_{A'}}^{A'}\!\exp(-i\zeta_\nu\theta')
 \,\frac{T_n(\theta'/A')\,d\theta'}{\sqrt{A'^2-\theta'^2}}
\end{eqnarray}
%%%%%%%%%%%%%%%%%%%%%%%%%%%%%%%%%%%%%%%%%%%%%%%%%%%%%%%%%%%%%%%   19
where $T_m(x)=\cos(m\arccos x)$~are Chebyshev polynomials,
$$
\theta_{A'}=\theta\;\;{\rm at}\;\,A^{'2}>\theta^2,\qquad{\rm and}\qquad
\theta_{A'}=A'\times{\rm sign}(\theta)\;\;\,{\rm at}\;\;A^{'2}<\theta^2.
$$
The assumption  $\zeta_\nu=0$ is used to start with 
but chromaticity will be added hereafter.

As it was established in previous section, the beam instability is governed
mostly by the multipoles $m=0,\;\pm1$, whereas other ones are of 
a minor importance.
Corresponding kernels are:
%
%%%%%%%%%%%%%%%%%%%%%%%%%%%%%%%%%%%%%%%%%%%%%%%%%%%%%%%%%%%%%%%   20
\begin{equation}
 K_{m,m}(A,A')=K_{m,-m}(A,A')=\delta_{m,0}
\end{equation}
%%%%%%%%%%%%%%%%%%%%%%%%%%%%%%%%%%%%%%%%%%%%%%%%%%%%%%%%%%%%%%%   20
%
%%%%%%%%%%%%%%%%%%%%%%%%%%%%%%%%%%%%%%%%%%%%%%%%%%%%%%%%%%%%%%%   21
\begin{eqnarray}
 K_{0,\pm1}(A,A')=-K_{\pm1,0}(A',A)                    
 = \frac{2}{\pi^2A'}\int_{-A}^A\sqrt{\frac{A'^2-\theta_{A'}^2}
 {A^2-\theta^2}}\,d\theta
\end{eqnarray}
%%%%%%%%%%%%%%%%%%%%%%%%%%%%%%%%%%%%%%%%%%%%%%%%%%%%%%%%%%%%%%%   21
%
(note that Eq.~(20) is valid with any $m$ at $\zeta_\nu=0$).
Therefore equations for three mentioned multipoles can be written 
in the form
%
%%%%%%%%%%%%%%%%%%%%%%%%%%%%%%%%%%%%%%%%%%%%%%%%%%%%%%%%%%%%%%%   22
\begin{eqnarray}
 \nu Y_0(A) = 2\pi q_0\int_0^\infty\!\Big\{\,Y_0(A')+K_{0,1}(A,A') 
\big[Y_{-1}(A') +Y_1(A')\big]\Big\}F(A')A'\,dA'                 
\end{eqnarray}
%%%%%%%%%%%%%%%%%%%%%%%%%%%%%%%%%%%%%%%%%%%%%%%%%%%%%%%%%%%%%%%   22
%
%%%%%%%%%%%%%%%%%%%%%%%%%%%%%%%%%%%%%%%%%%%%%%%%%%%%%%%%%%%%%%%   23
\begin{eqnarray}
 (\nu\mp Q_s) Y_{\pm1}(A)                              
=\,-2\pi q_0\int_0^\infty K_{0,1}(A',A)Y_0(A')F(A')A'\,dA'\;
\end{eqnarray}
%%%%%%%%%%%%%%%%%%%%%%%%%%%%%%%%%%%%%%%%%%%%%%%%%%%%%%%%%%%%%%%   23
%
Excluding $Y_{\pm1}$ one can obtain the ordinary integral equation 
for the function $Y_0(A)$:
%
%%%%%%%%%%%%%%%%%%%%%%%%%%%%%%%%%%%%%%%%%%%%%%%%%%%%%%%%%%%%%%%   24
\begin{eqnarray}
 \nu Y_0(A) = 2\pi q_0\int_0^\infty Y_0(A')F(A')A'\,dA'   
-\frac{8\pi^2q_0^2\nu}{\nu^2-Q_s^2}
 \int_0^\infty {\bf K}(A,A')Y_0(A')F(A')A'\,dA'
\end{eqnarray}
%%%%%%%%%%%%%%%%%%%%%%%%%%%%%%%%%%%%%%%%%%%%%%%%%%%%%%%%%%%%%%%   24
%
with the kernel
%
%%%%%%%%%%%%%%%%%%%%%%%%%%%%%%%%%%%%%%%%%%%%%%%%%%%%%%%%%%%%%%%   25
\begin{equation}
 {\bf K}(A,A') = \int_0^\infty\!\! K_{0,1}(A,X)K_{0,1}(A',X)F(X)\,XdX 
\end{equation}
%%%%%%%%%%%%%%%%%%%%%%%%%%%%%%%%%%%%%%%%%%%%%%%%%%%%%%%%%%%%%%%   25
%
Multiplying Eq.~(24) by $\,Y_0(A)F(A)A\,$ and integrating over $A$, 
one can get the simple equation for the eigentunes $\nu$ 
%
%%%%%%%%%%%%%%%%%%%%%%%%%%%%%%%%%%%%%%%%%%%%%%%%%%%%%%%%%%%%%%%   26
\begin{equation}
 \frac{\nu-\bar q}{2}\biggl(\nu-\frac{Q_s^2}{\nu}\biggr)=-\alpha^2 q_0^2
\end{equation}
%%%%%%%%%%%%%%%%%%%%%%%%%%%%%%%%%%%%%%%%%%%%%%%%%%%%%%%%%%%%%%%   26
%
with the parameters which do not depend on $\,\nu$:
%
%%%%%%%%%%%%%%%%%%%%%%%%%%%%%%%%%%%%%%%%%%%%%%%%%%%%%%%%%%%%%%%   27
\begin{equation}
 \bar q = \frac{2\pi q_0\,[\int_0^\infty Y_0(A)F(A)\,AdA]^2} 
{\int_0^\infty Y_0^2(A)F(A)A\,dA}
\end{equation}
%%%%%%%%%%%%%%%%%%%%%%%%%%%%%%%%%%%%%%%%%%%%%%%%%%%%%%%%%%%%%%%   27
%
and
%
%%%%%%%%%%%%%%%%%%%%%%%%%%%%%%%%%%%%%%%%%%%%%%%%%%%%%%%%%%%%%%%   28
\begin{equation}
 \alpha^2=\frac{4\pi^2\int_0^\infty\int_0^\infty {\bf K}(A,A')\,Y_0(A)Y_0(A')\,
 F(A)F(A')\,AA'\,dAdA'} {\int_0^\infty Y_0^2(A)F(A)A\,dA}
\end{equation}
%%%%%%%%%%%%%%%%%%%%%%%%%%%%%%%%%%%%%%%%%%%%%%%%%%%%%%%%%%%%%%%   28
%
These parameters are not too much sensitive to choice of the 
function $\,Y_0(A)$, so that some approximate solution of Eq.~(24)
can be used for the estimation.
The function $\,Y_0=1\,$ is a simple and rather reasonable option
because (i) it satisfies Eq.~(24) without coupling presenting the 
lower radial mode at $q_0\ll 1$, 
(ii) and it is one of the exact solutions of Eq.~(12) 
for hollow bunch with any $q_0$.
True, higher radial modes are excluded from consideration by this choice.
However, it does not matter in the case because it is known that 
these modes are much more stable \cite{SAH,HAND}.

With this choice,  the relation $\,\bar q=q_0\,$ follows from Eq.~(27),
and Eq.~(28) also obtains a compact form (see Appendix):
%
%%%%%%%%%%%%%%%%%%%%%%%%%%%%%%%%%%%%%%%%%%%%%%%%%%%%%%%%%%%%%%%   29
\begin{equation}
 \alpha^2 = \frac{8}{\pi}\int_0^\infty\!\!\! F(A)A^3\,dA\;
 \Big[\int_0^\pi\!\!\rho(A\cos\phi)\sin^2\phi\,d\phi\,\Big]^2
\end{equation}
%%%%%%%%%%%%%%%%%%%%%%%%%%%%%%%%%%%%%%%%%%%%%%%%%%%%%%%%%%%%%%%   29
%
%%%%%%%%%%%%%%%%%%%%%%%%%%%%%%%%%%%%%%%%%%%%%%%%%%%%%%%%%%%%%%%   Table 2
\begin{table}[b!]
\begin{center}
\caption{Parameters $\,\alpha\,$ and TMCI threshold of different 
bunch shapes}
\vspace{5mm}
\begin{tabular}{|c|c|c|c|c|c|}
\hline 
           & Hollow* & Hollow   & Rectang & Parabol & Gauss  \\
\hline
$\alpha  $ &   N/A   & 0.4053 &  0.4082   &  0.4075 &0.4020  \\
\hline
$|q_0|_{\rm threshold}/Q_s$
           &  0.5671 &  0.5689  & 0.5672  & 0.5676  &0.5708  \\
\hline
\end{tabular}
\end{center}
\end{table}
%%%%%%%%%%%%%%%%%%%%%%%%%%%%%%%%%%%%%%%%%%%%%%%%%%%%%%%%%%%%%%    Table 2
%
Results of calculated by this formula are collected in Table~II
for several distributions.
Somewhat surprising fact is very slight dependence of the parameters
on the bunch shape, even for so far apart models as the hollow and 
the Gaussian bunches.
It is confirmed by Fig.~2 where solutions of Eq.~(26) are plotted against 
the wake strength, and the images of different distributions are also 
indistinguishable.
The data taken from Section III are also transferred in this graph being 
presented by the dark circles, and demonstrating absolute agreement of 
the results obtained by so different methods. 
%
%%%%%%%%%%%%%%%%%%%%%%%%%%%%%%%%%%%%%%%%%%%%%%%%%%%%%%%%%%%%%%%   Fig.2
\begin{figure}[t!]
\includegraphics[width=85mm]{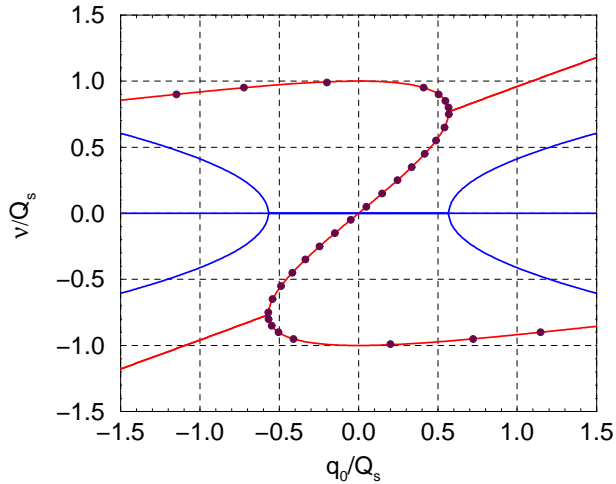}
\caption{Eigentunes of hollow bunch without chromaticity (3-modes 
approximation). 
Red and blue curves present real and imaginary parts of the tune. 
Solid lines are obtained by solution of Eq.~(25), 
and some data are transfered from Fig.~1 being presented by circles.}
\end{figure}
%%%%%%%%%%%%%%%%%%%%%%%%%%%%%%%%%%%%%%%%%%%%%%%%%%%%%%%%%%%%%%%   Fig.2
%

The eigentunes shown in Fig.~2 are real numbers at low wake.
However, they obtain an imaginary addition (blue lines) at rather large 
$\,q_0,$ which is the TMCI threshold in the case.
Corresponding values are presented in the last line of Table II 
being about 0.57 in all the examples.
In agreement with definition given by Eq.~(7), it allows to 
represent the TMCI threshold in usual terms 
%
%%%%%%%%%%%%%%%%%%%%%%%%%%%%%%%%%%%%%%%%%%%%%%%%%%%%%%%%%%%%%%%   30
\begin{equation}
 \frac{r_0RN|W_1|}{8\beta\gamma Q_0Q_s} > 0.57\pi\simeq 1.8
\end{equation}
%%%%%%%%%%%%%%%%%%%%%%%%%%%%%%%%%%%%%%%%%%%%%%%%%%%%%%%%%%%%%%%   30
%
which expression almost does not depend the bunch shape.
The result coincides very well with the known expressions
\cite{KOH,CHAO,CHIN,NG,HAND}. 

%
%%%%%%%%%%%%%%%%%%%%%%%%%%%%%%%%%%%%%%%%%%%%%%%%%%%%%%%%%%%%%%%   Sec. 5

\section{Three-mode model with chromaticity}

%%%%%%%%%%%%%%%%%%%%%%%%%%%%%%%%%%%%%%%%%%%%%%%%%%%%%%%%%%%%%%%   Sec. 5
%

As it has been shown in previous section, the function $\,Y_0=1\,$
is rather good approximation to describe the single bunch instability 
near threshold at zero chromaticity.
Besides, the relation
%
%%%%%%%%%%%%%%%%%%%%%%%%%%%%%%%%%%%%%%%%%%%%%%%%%%%%%%%%%%%%%%%   31
\begin{equation}
 Y_{\pm1}(A)\propto A \int_0^\pi \rho(A\cos\phi)\sin^2\phi\,d\phi 
\end{equation}
%%%%%%%%%%%%%%%%%%%%%%%%%%%%%%%%%%%%%%%%%%%%%%%%%%%%%%%%%%%%%%%   31
%
follows from Eq.~(21) and Eq.~(23) in this approximation.
Important thing is that the integral in this expression moderately depends 
on the amplitude at realistic distributions.
For example, it does not depend at all for the rectangular bunch, 
and has a variation not more of 25\% for the parabolic one.
Therefore the approximation $Y_{\pm 1}\propto A$ looks rather reasonably
for this case.
It means that all above presented results could be obtained using the 
pattern solution 
%
%%%%%%%%%%%%%%%%%%%%%%%%%%%%%%%%%%%%%%%%%%%%%%%%%%%%%%%%%%%%%%%   32
\begin{equation}
  Y(A,\phi) = 1+C_\theta\theta+C_uu
\end{equation}
%%%%%%%%%%%%%%%%%%%%%%%%%%%%%%%%%%%%%%%%%%%%%%%%%%%%%%%%%%%%%%%   32
%
with indefinite constants $C_\theta$ and $C_u$.
Confirmations of this statement will be furnished later.
However, the main thing is that this model paves the way 
to extend the theory by including chromaticity, space charge, etc. 
Chromaticity is the fist point which will be applied in this section.

Substitution of Eq.~(32) to Eq.~(8) results in 
%
%%%%%%%%%%%%%%%%%%%%%%%%%%%%%%%%%%%%%%%%%%%%%%%%%%%%%%%%%%%%%%%   33
\begin{eqnarray}
 \nu+(\nu C_\theta\!+\!iQ_sC_u)\,\theta+(\nu C_u\!-\!iQ_sC_\theta)\,u 
 = 2q_0\exp(i\zeta\theta)\int_\theta^\infty \!\!\!\!\rho(\theta')
 (1\!+\!C_\theta\theta')\exp(-i\zeta\theta')\,d\theta'\quad\;\;
\end{eqnarray}
%%%%%%%%%%%%%%%%%%%%%%%%%%%%%%%%%%%%%%%%%%%%%%%%%%%%%%%%%%%%%%%   33
%
The relation $\,C_u=iC_\theta Q_s/\nu\,$ follows from this immediately. 
Two more equations can be obtained by multiplication of Eq.~(33) 
by $\rho(\theta)$ or $\rho(\theta)\theta$ with subsequent 
integration over $\;\theta$.
Then, excluding parameter $C_\theta$, one can get required dispersion 
equation for the eigentunes $\nu$.
It is represented below for the case $|\zeta\theta_0|<\sim0.5$ 
which assumption allows to estimate effect of chromaticity 
without excessively bulky expressions:
%
%%%%%%%%%%%%%%%%%%%%%%%%%%%%%%%%%%%%%%%%%%%%%%%%%%%%%%%%%%%%%%%   34
\begin{equation}
 \frac{\nu-q_0(1-i\alpha\chi)}{2}\left(\nu-\frac{Q_s^2}{\nu}
 -2i\beta q_0\chi\right)\simeq-q_0^2\left(\alpha -\frac{i\chi}{4}\right)^2
\end{equation}
%%%%%%%%%%%%%%%%%%%%%%%%%%%%%%%%%%%%%%%%%%%%%%%%%%%%%%%%%%%%%%%   34
%
where $\chi=2\sqrt2\,\zeta_\nu\sigma_\theta$, and following designations
are applied: 
%
%%%%%%%%%%%%%%%%%%%%%%%%%%%%%%%%%%%%%%%%%%%%%%%%%%%%%%%%%%%%%%%   35
\begin{subequations}
\begin{eqnarray}
 \sigma_\theta^2  = \int_{-\infty}^\infty\rho(\theta) \,\theta^2d\theta,\; 
 \hspace{31mm}  
 \end{eqnarray}
%%%%%%%%%%%%%%%%%%%%%%%%%%%%%%%%%%%%%%%%%%%%%%%%%%%%%%%%%%%%%%%   35a
 \begin{eqnarray}
 \alpha=\frac{\sqrt{2}}{\sigma_\theta}\int_{-\infty}^\infty\rho(\theta)\,
 \theta d\theta \int_0^\theta\rho(\theta')\,d\theta',\quad\;
 \end{eqnarray}
%%%%%%%%%%%%%%%%%%%%%%%%%%%%%%%%%%%%%%%%%%%%%%%%%%%%%%%%%%%%%%%   35b
 \begin{eqnarray}
 \beta=\frac{1}{\sigma_\theta^3\sqrt{2}}\int_{-1}^1\rho(\theta)\,\theta 
 d\theta \int_0^\theta \rho(\theta')\,\theta'^2d\theta'
\end{eqnarray}
\end{subequations}
%%%%%%%%%%%%%%%%%%%%%%%%%%%%%%%%%%%%%%%%%%%%%%%%%%%%%%%%%%%%%%%   35c
%
%%%%%%%%%%%%%%%%%%%%%%%%%%%%%%%%%%%%%%%%%%%%%%%%%%%%%%%%%%%%%%%   Table 3
\begin{table}[b!]
\begin{center}
\caption{Dispersion equation parameters of different bunches}
\vspace{5mm}
\begin{tabular}{|c|c|c|c|c|c|c|}
\hline 
                 & Hollow      & Boxcar      & Parabolic      & Gaussian \\
\hline
$\quad\sigma_\theta^2\quad$&$\qquad\theta_0^2/2\qquad$&
$\quad\theta_0^2/3\qquad$
&$\qquad\theta_0^2/5\qquad$& Any   \\
$\alpha$         & 0.405       & 0.408       & 0.407          & 0.400 \\
$\beta $         & 0.135       & 0.123       & 0.113          & 0.100 \\
\hline
\end{tabular}
\end{center}
\end{table}
%%%%%%%%%%%%%%%%%%%%%%%%%%%%%%%%%%%%%%%%%%%%%%%%%%%%%%%%%%%%%%    Table 3
%
%%%%%%%%%%%%%%%%%%%%%%%%%%%%%%%%%%%%%%%%%%%%%%%%%%%%%%%%%%%%%%%%   Fig.3
\begin{figure*}[b!]
\hspace{-17mm}
\begin{minipage}[b!]{0.45\linewidth}
\begin{center}
\includegraphics[width=85mm]{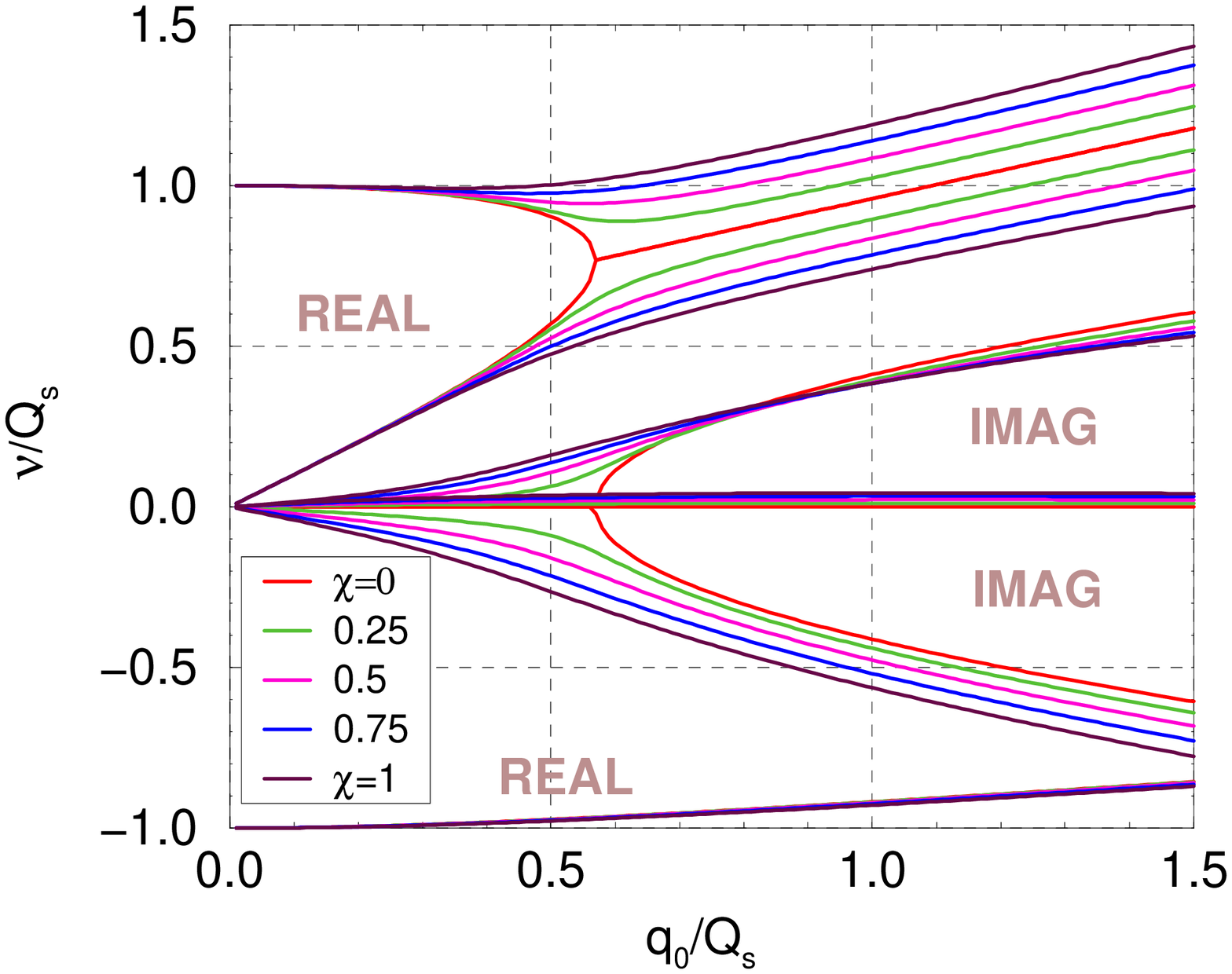}
\end{center}
\end{minipage}
\hspace{-7mm}
\begin{minipage}[b!]{0.45\linewidth}
\begin{center}
\includegraphics[width=85mm]{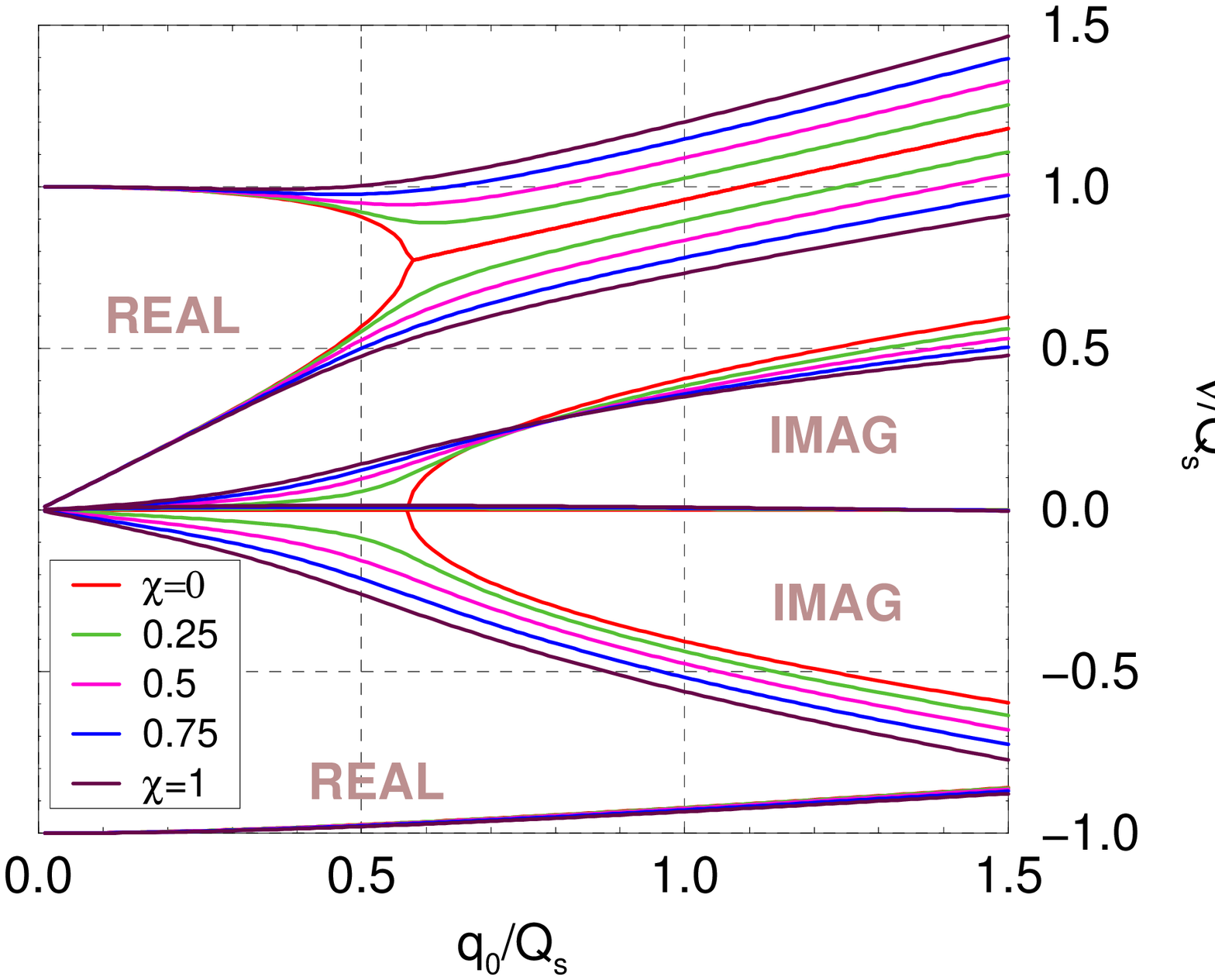}
\end{center}
\end{minipage}
\caption{Eigentunes with chromaticity: left -- hollow bunch, right -- 
Gaussian distribution.
All the curves are odd functions of the wake strength.
Real parts do not depend on sign of chromaticity, imaginary parts 
change sign.  
Some curves are indistinguishable because the modes starting from 
the point $\nu=-Q_s$ almost do not depend on chromaticity.}
\end{figure*}
%%%%%%%%%%%%%%%%%%%%%%%%%%%%%%%%%%%%%%%%%%%%%%%%%%%%%%%%%%%%%%%   Fig.3
%
Because $\sigma_\theta$ is rms bunch length, $\chi$~can be treated 
as betatron phase advance caused by chromaticity in the entire bunch
(it is really true for the hollow bunch of length $\,2\theta_0\,$
when $\,\sigma_\theta=\theta_0/\sqrt2$).
Other parameters are represented in Table~III for several distributions.
Comparison with Table II let us to conclude that Eq.~(26) and (34) coincide 
not only formally but also actually at $\chi=0$, 
because the difference of coefficients $\alpha$ is negligible.
It can be concluded as well that the dependence of the eigentunes on the 
bunch shape is very weak when the chromaticity is also included. 
The statement is confirmed by Fig.~3 where complex solutions of Eq.~(34) 
are plotted against the wake strength at different chromaticity, 
for hollow and Gaussian bunches.
Note that only positive wakes and chromaticities are presented in the graphs 
because all curves have following symmetry properties:
(i)~they are odd functions of $q_0/Q_s$; 
(ii)~real parts of the tunes do not depend on sign of chromaticity; 
(iii)~the imaginary additions reflect specularly with respect to the line 
$\nu=0$, when the chromaticity change sign. 
It is seen that the instability has no threshold with chromaticity,
and TMCI appears against the head-tail background without a pronounced
demarcation line. 
In particular, the head-tail and TMCI contributions are comparable 
at $\,|q_0|\simeq 0.7\,Q_s\,$ and $\,|\theta_0\zeta|\simeq 0.5$. 
It should be noted in addition that no sign of chromaticity can prevent 
instability of all bunch modes. 

With accuracy of several percents, solutions of Eq.~(34) at $\,q_0<Q_s\,$ 
can be presented in the form
%
%%%%%%%%%%%%%%%%%%%%%%%%%%%%%%%%%%%%%%%%%%%%%%%%%%%%%%%%%%%%%%%   36
\begin{eqnarray}
 \nu\;\simeq\;\frac{q_0+mQ_s-iq_0\chi(\alpha-\beta)}{2}    
 \pm\sqrt{\biggl(\frac{q_0-mQ_s-iq_0\chi(\alpha+\beta)}{2}\biggr)^2
 -q_0^2\biggl(\alpha-\frac{i\chi}{4}\biggr)^2    }
\end{eqnarray}
%%%%%%%%%%%%%%%%%%%%%%%%%%%%%%%%%%%%%%%%%%%%%%%%%%%%%%%%%%%%%%%   36
%
where $\,m=\pm1$.
In particular, it provides correct TMCI threshold without chromaticity, 
and leads to well known formulae for the head-tail modes at $\,q_0\ll Q_s$.
In the last case the solutions for hollow bunch can be reduced 
to the form: 
%
%%%%%%%%%%%%%%%%%%%%%%%%%%%%%%%%%%%%%%%%%%%%%%%%%%%%%%%%%%%%%%%   37
\begin{equation}
 \nu_m=mQ_s+q_0\delta_{m,0}+\frac{8iq_0\theta_0\zeta_\nu}{\pi^2(4m^2-1)}
\end{equation}
%%%%%%%%%%%%%%%%%%%%%%%%%%%%%%%%%%%%%%%%%%%%%%%%%%%%%%%%%%%%%%%   37
%
which expression is valid with any $\,m\,$ \cite{CHAO}. 
Analyzing Table III, one can add that this result almost does not 
depend on the bunch shape, at least for lowest radial modes and multipoles.

%
%%%%%%%%%%%%%%%%%%%%%%%%%%%%%%%%%%%%%%%%%%%%%%%%%%%%%%%%%%%%%%%   SEC.6
%
\section{Realistic bunch with arbitrary wake}
%
%%%%%%%%%%%%%%%%%%%%%%%%%%%%%%%%%%%%%%%%%%%%%%%%%%%%%%%%%%%%%%%   SEC.6
%

It would be  beyond reasons to treat rectangular wake as 
an exclusive case. 
On the contrary, Eq.~(32) can be applied as an approximate solution of 
general Eq.~(6) to look for the eigentunes of a bunch with arbitrary wake 
function. 
Subsequent transformations are described just after Eq.~(32) and result 
in the dispersion equation like Eq.~(34) or (26):
%
%%%%%%%%%%%%%%%%%%%%%%%%%%%%%%%%%%%%%%%%%%%%%%%%%%%%%%%%%%%%%%%   38
\begin{equation}
 \frac{\nu-q_{ef}}{2}\left(\nu-\frac{Q_s^2}{\nu}-2\alpha_2 q_{ef}\right) 
=-\alpha_1^2 q_{ef}^2
\end{equation}
%%%%%%%%%%%%%%%%%%%%%%%%%%%%%%%%%%%%%%%%%%%%%%%%%%%%%%%%%%%%%%%   38
%
with the coefficients 
%
%%%%%%%%%%%%%%%%%%%%%%%%%%%%%%%%%%%%%%%%%%%%%%%%%%%%%%%%%%%%%%%   39
\begin{subequations}
\begin{eqnarray}
 q_{ef} = 2\int_0^\infty \tilde q(\theta)d\theta\;     
 \int_{-\infty}^{\infty}\rho(\theta'-\theta/2)\;
 \rho(\theta'+\theta/2)\,d\theta'\hspace{27mm}                             
\end{eqnarray}
%%%%%%%%%%%%%%%%%%%%%%%%%%%%%%%%%%%%%%%%%%%%%%%%%%%%%%%%%%%%%%%   39a
\begin{eqnarray}
 \alpha_1 = \frac{1}{\sigma_\theta q_{ef}\sqrt2}\int_0^\infty 
 \tilde q(\theta)\,\theta d\theta 
 \times\int_{-\infty}^{\infty}
 \rho(\theta'-\theta/2)\,\rho(\theta'+\theta/2)\,d\theta'\qquad      
\end{eqnarray}
%%%%%%%%%%%%%%%%%%%%%%%%%%%%%%%%%%%%%%%%%%%%%%%%%%%%%%%%%%%%%%%   39b
\begin{eqnarray}
 \alpha_2=\frac{1}{\sigma_\theta^2q_{ef}}\int_0^\infty \tilde q(\theta)
 \,d\theta\int_{-\infty}^\infty \!\!\rho(\theta'-\theta/2)\,
 \rho(\theta'+\theta/2)\,(\theta'^2-\theta^2/4)\,d\theta'  
\end{eqnarray}
\end{subequations}
%%%%%%%%%%%%%%%%%%%%%%%%%%%%%%%%%%%%%%%%%%%%%%%%%%%%%%%%%%%%%%%   39c
%
where $\sigma_\theta$ is rms bunch length given by Eq.~(35a),
and $\,\tilde q(\theta)=q(\theta)\exp(-i\zeta_\nu\theta)$.
Approximate solutions of Eq.~(38) can be presented in the form
like Eq.~(36):
%
%%%%%%%%%%%%%%%%%%%%%%%%%%%%%%%%%%%%%%%%%%%%%%%%%%%%%%%%%%%%%%%   40
\begin{equation}
 \nu \simeq \frac{q_{ef}(1+\alpha_2)+mQ_s}{2}\pm\sqrt{\biggl[
 \frac{q_{ef}(1-\alpha_2)-mQ_s}{2}\biggr]^2\!\!\!-\alpha_1^2 q_{ef}^2 }
\end{equation}
%%%%%%%%%%%%%%%%%%%%%%%%%%%%%%%%%%%%%%%%%%%%%%%%%%%%%%%%%%%%%%%   40
%
with $\,m=\pm1$.
Although parameter $\,\chi\,$ does not appear in the expression,
chromaticity is still presented here being included in 
the functions $\,\tilde q(\theta)$ and $q_{\rm ef}$.
However, next consideration will be restricted by the case 
of zero chromaticity: $\,\zeta_\nu=0,\;\tilde q(\theta)=q(\theta)$. 
Then $\alpha_{1,2}$ are real numbers, and the instability can appear 
only in the TMCI form with the threshold: 
%
%%%%%%%%%%%%%%%%%%%%%%%%%%%%%%%%%%%%%%%%%%%%%%%%%%%%%%%%%%%%%%%   41
\begin{equation}
 |q_{ef}|_{thresh}=\frac{Q_s}{1+2\alpha},\qquad 
 \alpha=\alpha_1-\frac{\alpha_2}{2} 
\end{equation}
%%%%%%%%%%%%%%%%%%%%%%%%%%%%%%%%%%%%%%%%%%%%%%%%%%%%%%%%%%%%%%%   41
%
For Gaussian bunch with dispersion $\sigma_\theta$, used parameters 
obtain the forms
%
%%%%%%%%%%%%%%%%%%%%%%%%%%%%%%%%%%%%%%%%%%%%%%%%%%%%%%%%%%%%%%%   42
\begin{subequations}
\begin{equation}
 q_{ef} = \frac{1}{\sigma_\theta\sqrt\pi}\int_0^\infty\exp\biggl(-\frac
 {\theta^2}{4\sigma_\theta^2}\biggr)\,q(\theta)\,d\theta\hspace{28mm}         
\end{equation}
%%%%%%%%%%%%%%%%%%%%%%%%%%%%%%%%%%%%%%%%%%%%%%%%%%%%%%%%%%%%%%%   42a
\begin{equation}
 \alpha_1 = \frac{1}{2\sqrt{2\pi}\,\sigma_\theta^2q_{ef}}\int_0^\infty
 \exp\biggl(-\frac{\theta^2}{4\sigma_\theta^2}\biggr)\,q(\theta)
\,\theta d\theta \hspace{16mm}                                                 
\end{equation}
%%%%%%%%%%%%%%%%%%%%%%%%%%%%%%%%%%%%%%%%%%%%%%%%%%%%%%%%%%%%%%%   42b
\begin{equation}
 \alpha_2=\frac{1}{4\sqrt\pi\,\sigma_\theta q_{ef}}\int_0^\infty
 \biggl(1-\frac{\theta^2}{2\sigma_\theta^2}\biggr)
 \exp\biggl(-\frac{\theta^2}{4\sigma_\theta^2}\biggr)\,q(\theta)\,d\theta
\end{equation}
\end{subequations}
%%%%%%%%%%%%%%%%%%%%%%%%%%%%%%%%%%%%%%%%%%%%%%%%%%%%%%%%%%%%%%%   42c
%
Note that $\,q_{ef}=q_0,\,\alpha_1=\alpha,\,\alpha_2=0\,$ at 
constant wake $\,q=q_0$.
Two more examples are considered below.

%%%%%%%%%%%%%%%%%%%%%%%%%%%%%%%%%%%%%%%%%%%%%%%%%%%%%%%%%%%%%%%
\subsection
{Short rectangular wake}
%%%%%%%%%%%%%%%%%%%%%%%%%%%%%%%%%%%%%%%%%%%%%%%%%%%%%%%%%%%%%%%

%%%%%%%%%%%%%%%%%%%%%%%%%%%%%%%%%%%%%%%%%%%%%%%%%%%%%%%%%%%%%%%   Fig.4
\begin{figure}[t!]
\begin{center}
\includegraphics[width=85mm]{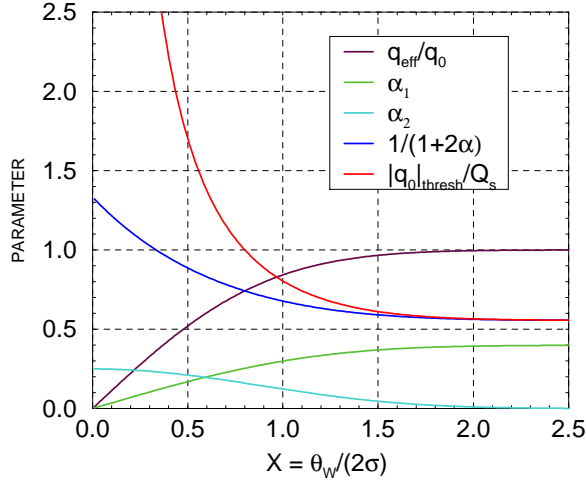}
\end{center}
\caption{Parameters of Gaussian bunch with short rectangular wake.
Brown line describes an effective weakening of the wake in 
a short bunch, red line represents TMCI threshold.}
\end{figure}
%%%%%%%%%%%%%%%%%%%%%%%%%%%%%%%%%%%%%%%%%%%%%%%%%%%%%%%%%%%%%%%   Fig.4
%
Gaussian bunch with a rectangular wake of restricted length $\theta_w$
is considered in this subsection as first example (similar wake can be 
created e.g. by a pair of strip-line BPM \cite{HAND}).
Eq.~(42) gives in this case
%
%%%%%%%%%%%%%%%%%%%%%%%%%%%%%%%%%%%%%%%%%%%%%%%%%%%%%%%%%%%%%%%   43
\begin{equation}
 q_{ef} = q_0\,{\rm erf}(x),\qquad
 \alpha_1 = \frac{1-\exp(-x^2)}{\sqrt{2\pi}\,{\rm erf}\,(x)},\qquad
 \alpha_2=\frac{x\exp(-x^2)}{2\sqrt\pi\,{\rm erf}\,(x)}
\end{equation}
%%%%%%%%%%%%%%%%%%%%%%%%%%%%%%%%%%%%%%%%%%%%%%%%%%%%%%%%%%%%%%%   43
where $x=\theta_w/(2\sigma_\theta)$.
These functions are plotted in Fig.~4. 
Threshold value of $\,q_0\,$ is shown as well being determined with 
help of the expression
%
%%%%%%%%%%%%%%%%%%%%%%%%%%%%%%%%%%%%%%%%%%%%%%%%%%%%%%%%%%%%%%%   44
\begin{equation}
 \frac{|q_0|_{thresh}}{Q_s}=\frac{1}{[1+2\alpha(x)]\,{\rm erf}\,(x)}
\end{equation}
%%%%%%%%%%%%%%%%%%%%%%%%%%%%%%%%%%%%%%%%%%%%%%%%%%%%%%%%%%%%%%%   44
%
It is seen that a shortening of the wake results in a rise of the 
threshold which becomes especially noticeable at $\,\theta_w<\sim 2\sigma\,$.  

%
%%%%%%%%%%%%%%%%%%%%%%%%%%%%%%%%%%%%%%%%%%%%%%%%%%%%%%%%%%%%%%%
\subsection {Resistive wake}
%%%%%%%%%%%%%%%%%%%%%%%%%%%%%%%%%%%%%%%%%%%%%%%%%%%%%%%%%%%%%%%
%

Resistive wall impedance is the most general and 
important source of transverse instabilities in circular accelerators.
Corresponding normalized wake function is 
%
%%%%%%%%%%%%%%%%%%%%%%%%%%%%%%%%%%%%%%%%%%%%%%%%%%%%%%%%%%%%%%%   45
\begin{equation}
 q(\theta) = \frac{q_{rw}}{\sqrt\theta},\qquad\quad
{q_{rw}}=-\frac{r_0R^2N}{2\pi\beta\gamma Q_0b^3}
\sqrt{\frac{c}{R\sigma_c}}
\end{equation}
%%%%%%%%%%%%%%%%%%%%%%%%%%%%%%%%%%%%%%%%%%%%%%%%%%%%%%%%%%%%%%%   45
%
where $b$ is the beam pipe radius, and $\sigma_c$ is the pipe wall
conductivity (see e.g. \cite{HAND}).
With this wake, integrals in Eq.~(42) are representable
in terms of gamma functions:  
%
%%%%%%%%%%%%%%%%%%%%%%%%%%%%%%%%%%%%%%%%%%%%%%%%%%%%%%%%%%%%%%%   46
\begin{eqnarray}
 q_{ef}=\frac{q_{rw}\,\Gamma(1/4)} {\sqrt{2\pi\sigma_\theta}}
=\frac{1.4464\,q_{rw}}{\sqrt{\sigma_\theta}},\qquad          
 \alpha_1=\frac{\Gamma(3/4)}{\sqrt2\,\Gamma(1/4)}=0.2390,\qquad      
 \alpha_2=\frac{1}{8}
\end{eqnarray}
%%%%%%%%%%%%%%%%%%%%%%%%%%%%%%%%%%%%%%%%%%%%%%%%%%%%%%%%%%%%%%%   46
%
Threshold value of the effective and usual wakes can be found then 
with help of Eq.~(41): $\;|q_{ef}|_{thresh}=0.739\,Q_s,\;\;
|q_{rw}|_{thresh}=0.511\,Q_s\sqrt\sigma_\theta.\;$
Therefore the resistive wall TMCI threshold in usual terms is 
%
%%%%%%%%%%%%%%%%%%%%%%%%%%%%%%%%%%%%%%%%%%%%%%%%%%%%%%%%%%%%%%%   47
\begin{equation}
 \frac{r_0R^2N_{thresh}}{2\pi\beta\gamma Q_0Q_sb^3}
 \sqrt{\frac{c}{\sigma_c\sigma_z}} = 0.51
\end{equation}
%%%%%%%%%%%%%%%%%%%%%%%%%%%%%%%%%%%%%%%%%%%%%%%%%%%%%%%%%%%%%%%   47
%
where standard rms bunch length $\,\sigma_z=\sigma_\theta R\,$ is used.
However, it is necessary to take into account that Eq.~(45) is valid only at 
$R\theta>\sim b/\gamma$ when the wake reaches a maximum.     
Therefore sufficient condition of applicability of Eq.~(47) is
$\sigma_z\gg b/\gamma$. 

Another restriction comes from the fact that the resistive wake  
has a long and slowly decaying tail.
Therefore it can impact not only next bunches but also itself 
by the succeeding turns.    
These multibunch/multiturn collective effects should be included in 
a comprehensive investigation of resistive wall instability. 
However, this point is beyond the scope of the paper where only single 
bunch effects are examined.
Nevertheless it can be noted that presented results give a possibility 
to estimate a relative danger of the effects by a comparison of the 
contributed tune shifts.
Indeed, TMCI of a single bunch appears at 
$\,|\nu|_{\rm TMCI}\simeq 0.77\,Q_s\,$ as it follows from Fig.~1-3,
and corresponding bunch population is determined by Eq.~(47).
The collective modes i.e. long term tune shift with this intensity is \cite{B2}
%
%%%%%%%%%%%%%%%%%%%%%%%%%%%%%%%%%%%%%%%%%%%%%%%%%%%%%%%%%%%%%%%   48
\begin{equation}
 |\nu|_{\rm LONG}\simeq 0.51\,Q_s \sqrt{\frac{2\beta\sigma_z}{2\pi R|k-Q_0|}}
 \biggr(h-\frac{(2|k-Q_0|)^{3/2}}{h^{1/2}}\biggr)
\end{equation}
%%%%%%%%%%%%%%%%%%%%%%%%%%%%%%%%%%%%%%%%%%%%%%%%%%%%%%%%%%%%%%%   48
%
where $h$ is number of bunches, and $k$ is the collective mode number 
(the mode can be unstable at $k>Q_0$). 
Taking $\,|k-Q_0|=0.25\,$ and $\,\beta=1\,$ as a typical example,
we can compare these long term and TMCI effects as the ratio of 
corresponding tune shifts:
%
%%%%%%%%%%%%%%%%%%%%%%%%%%%%%%%%%%%%%%%%%%%%%%%%%%%%%%%%%%%%%%%   49
\begin{equation}
 \biggl|\frac{\nu_{\rm LONG}}{\nu_{TMCI}}\biggr| \sim
 2\,\biggr(h-\frac{0.35}{\sqrt h}\biggr)\sqrt{\frac{\sigma_z}{2\pi R}}
\end{equation}
%%%%%%%%%%%%%%%%%%%%%%%%%%%%%%%%%%%%%%%%%%%%%%%%%%%%%%%%%%%%%%%   49
%
With great probability this value is $\,<1\,$ or even $\,\ll1\,$ at $\,h=1$, 
that is the multiturn effect of a single bunch is typically small or even 
negligible in comparison with TMCI or head-tail instability. 
However, the collective effects are more dangerous in a multibunch 
machine with $\,h\gg1,\;h\sigma_z\sim R$.
Of course, more detailed analysis is required at intermediate cases.
  
%
%%%%%%%%%%%%%%%%%%%%%%%%%%%%%%%%%%%%%%%%%%%%%%%%%%%%%%%%%%%%%%%   SEC 6
%
\section
{Space charge effects}
%
%%%%%%%%%%%%%%%%%%%%%%%%%%%%%%%%%%%%%%%%%%%%%%%%%%%%%%%%%%%%%%%   SEC 6
%
%
%%%%%%%%%%%%%%%%%%%%%%%%%%%%%%%%%%%%%%%%%%%%%%%%%%%%%%%%%%%%%%%   Fig.5
\begin{figure}[t!]
\begin{center}
\includegraphics[width=85mm]{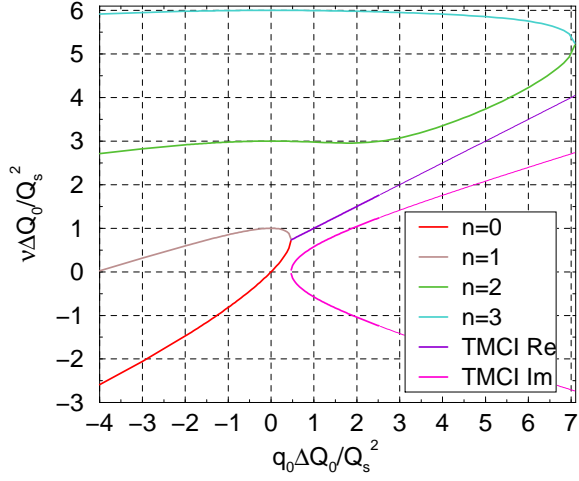}
\end{center}
\caption{Eigentunes of a rectangular (``boxcar'') bunch 
 with ultimate space charge $\Delta Q_0\gg Q_s$. 
 The graph is taken from \cite{B1};  $n = m(m+1)/2$.}
\end{figure}
%%%%%%%%%%%%%%%%%%%%%%%%%%%%%%%%%%%%%%%%%%%%%%%%%%%%%%%%%%%%%%%   Fig.5
%
Bunched beam instability with extremely large space charge 
$(\Delta Q\gg Q_s)$ was considered in works \cite{BU,B1,B2}.
The most remarkable phenomenon is a pronounced asymmetry of the curves
with respect to the wake sign which effect has been first shown in 
Ref.~\cite{B1}.
It is illustrated by Fig.~5 taken from the quoted article where
a rectangular (``boxcar'') bunch with constant wake was explored 
at space charge betatron tune shift $\Delta Q_0\gg Q_s$.
It is seen that TMCI appears only with the positive wake satisfying the
instability condition 
%
%%%%%%%%%%%%%%%%%%%%%%%%%%%%%%%%%%%%%%%%%%%%%%%%%%%%%%%%%%%%%%%   50
\begin{equation}
 (q_0)_{thresh} \simeq \frac{0.5\,Q_s^2}{\Delta Q_0} 
\end{equation}
%%%%%%%%%%%%%%%%%%%%%%%%%%%%%%%%%%%%%%%%%%%%%%%%%%%%%%%%%%%%%%%   50
%
which drastically differs from the conditions presented by Fig.~1-3, 
and Eq.~(30) of this paper.
It is needless to say that investigation of the effect at 
$\Delta Q_0\sim Q_s$ is the only way to resolve the problem by a 
joining of these conflicting pictures.
It turns out that the 3-modes model described by Eq.~(32) bridges 
these ultimate cases providing a general form of the lowers eigenmodes
over a wide range of parameters.

It is easy to verify that substitution of Eq.~(32) to general Eq.~(6) 
results in
%
%%%%%%%%%%%%%%%%%%%%%%%%%%%%%%%%%%%%%%%%%%%%%%%%%%%%%%%%%%%%%%%   51
\begin{eqnarray}
 \nu+(\nu C_\theta\!+\!iQ_sC_u)\,\theta+
 \big\{\big[\,(\Delta Q_{av}\rho(\theta)+\nu\big] 
 C_u\!-\!iQ_sC_\theta]\big\}\,u                              \nonumber  \\
 =\:2q_0\exp(i\zeta\theta)\int_\theta^\infty\rho(\theta')
 (1+C_\theta\theta')\exp(-i\zeta\theta')\,d\theta'
\end{eqnarray}
%%%%%%%%%%%%%%%%%%%%%%%%%%%%%%%%%%%%%%%%%%%%%%%%%%%%%%%%%%%%%%%   51
%
The only distinction of this expression from Eq.~(33) without space charge
is the addition proportional to $\Delta Q_{av}\rho(\theta)$. 
We consider in this section a rectangular bunch which density does not 
depend on longitudinal coordinate so that $Q_{av}\rho$ is the constant 
value $\,\Delta Q_0\,$ which coincides with incoherent space charge tune 
shift averaged over all transverse coordinates \cite{B2}. 
Therefore relation between coefficients of Eq.~(51) obtains the form
$\,C_u=iC_\theta Q_s/(\nu+\Delta Q_0)\,$ instead of 
$\,C_u=iC_\theta Q_s/\nu$.
As a result, all subsequent relations hold true with the replacement of 
$\,Q_s^2/\nu\,$ on $\,Q_s^2/(\nu+\Delta Q_0)$.
In particular, Eq.~(34) with constant wake function and chromaticity
obtains the form  
%
%%%%%%%%%%%%%%%%%%%%%%%%%%%%%%%%%%%%%%%%%%%%%%%%%%%%%%%%%%%%%%%   52
\begin{equation}
 \frac{\nu-q_0(1-i\alpha\chi)}{2}\left(\nu-\frac{Q_s^2}{\nu+\Delta Q_0}
 -2i\beta q_0\chi\right)\simeq-q_0^2\left(\alpha -\frac{i\chi}{4}\right)^2
\end{equation}
%%%%%%%%%%%%%%%%%%%%%%%%%%%%%%%%%%%%%%%%%%%%%%%%%%%%%%%%%%%%%%%   52
%
In the ``head-tail'' limit, that is at $\,|q_0|\ll Q_s\,$, 
approximate solutions of the equation are
%
%%%%%%%%%%%%%%%%%%%%%%%%%%%%%%%%%%%%%%%%%%%%%%%%%%%%%%%%%%%%%%%   53
\begin{equation}
 \nu_0=q_0(1-i\alpha\chi),\qquad \nu_{\pm 1}\simeq \pm\sqrt
 {Q_s^2+\frac{\Delta Q_0^2}{4}}-\frac{\Delta Q_0}{2}+i\beta q_0\chi 
\end{equation}
%%%%%%%%%%%%%%%%%%%%%%%%%%%%%%%%%%%%%%%%%%%%%%%%%%%%%%%%%%%%%%%   53
%
Thus space charge does not affect zero mode at all and does not change 
growth rate of the modes $\,m=\pm1\,$ in this ``head-tail'' approximation. 

Another situation arises at $\,q_0\sim Q_s\,$ or $\,q_0> Q_s\,$
where TMCI can arise. 
This case is illustrated by Fig.~6 where the eigentunes are plotted against 
the wake strength at different space charge, but without chromaticity.
Effect of the wake sign is seen very clearly in this graph: 
space charge propels the TMCI threshold to the centerline 
at $q_0>0$, and away from it at $q_0<0$.
Corresponding dependence is shown quantitatively in Fig.~7 where the
thresholds are presented separately for positive and negative wakes 
and supplemented by appropriate analytical formulae.
There is very good agreement of these results with Fig.~5. 
For example, TMCI threshold of positive wake is
%
%%%%%%%%%%%%%%%%%%%%%%%%%%%%%%%%%%%%%%%%%%%%%%%%%%%%%%%%%%%%%%%  54
\begin{equation}
 (q_0)_{thresh}\simeq\frac{0.57\,Q_s^2}{Q_s+\Delta Q_0} 
 \rightarrow \frac{0.57\,Q_s^2}{\Delta Q_0}
\end{equation}
%%%%%%%%%%%%%%%%%%%%%%%%%%%%%%%%%%%%%%%%%%%%%%%%%%%%%%%%%%%%%%%  54
what is very close to the estimation given by Fig.~5 and Eq.~(50).

However, space charge tune shift raises the TMCI threshold of
negative wakes.
For example, Eq.~(47) for resistive wall TMCI threshold obtains 
the form:
%
%%%%%%%%%%%%%%%%%%%%%%%%%%%%%%%%%%%%%%%%%%%%%%%%%%%%%%%%%%%%%%%   55
\begin{equation}
 \frac{r_0R^2N_{thresh}}{2\pi\beta\gamma Q_0Q_sb^3}
 \sqrt{\frac{\Omega_0R}{\sigma_c\sigma_z}} \simeq
 0.51\biggl(1+\frac{0.7\Delta Q_0}{Q_s}+\frac{0.3\Delta Q_0^2}{Q_s^2}\biggr)
\end{equation}
%%%%%%%%%%%%%%%%%%%%%%%%%%%%%%%%%%%%%%%%%%%%%%%%%%%%%%%%%%%%%%%   55
%

Joint effect of space charge and chromaticity is illustrated by Fig.~8 
where dependence of the eigentunes of rectangular bunch on the wake 
strength is presented at $\Delta Q_0=Q_s$ and various chromaticity.  
%
%%%%%%%%%%%%%%%%%%%%%%%%%%%%%%%%%%%%%%%%%%%%%%%%%%%%%%%%%%%%%%%   Fig.6
\begin{figure}[t!]
\begin{center}
\includegraphics[width=85mm]{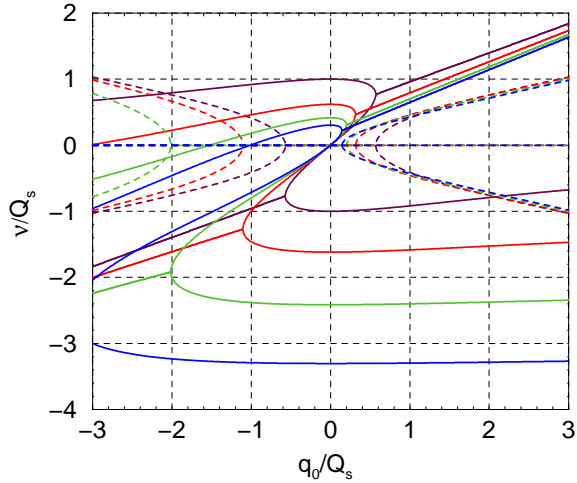}
\end{center}
\caption{Eigentunes of rectangular bunch with space charge
(chromaticity is turned off).  
Used ratios $\,\Delta Q_0/Q_s\,$ are: 0 (maroon), 1 (red), 2 (green), 
and 3 (blue). 
Solid/dashed lines represent real/imaginary parts of the tune.}
\end{figure}
%%%%%%%%%%%%%%%%%%%%%%%%%%%%%%%%%%%%%%%%%%%%%%%%%%%%%%%%%%%%%%%   Fig.6
%
%%%%%%%%%%%%%%%%%%%%%%%%%%%%%%%%%%%%%%%%%%%%%%%%%%%%%%%%%%%%%%%   Fig.7
\begin{figure}[h!]
\vspace{10mm}
\begin{center}
\includegraphics[width=85mm]{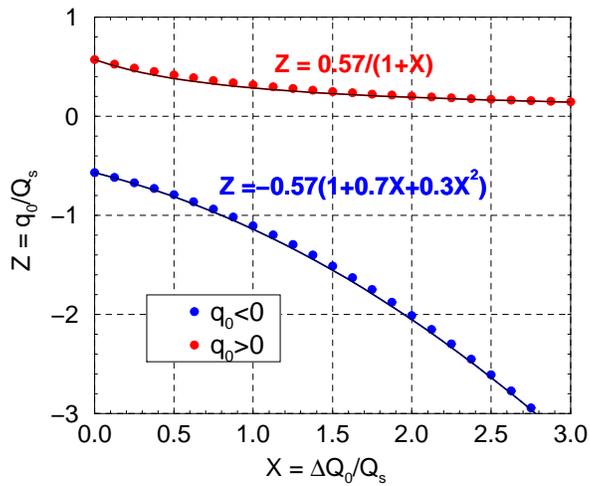}
\end{center}
\caption{TMCI threshold of the rectangular bunch against 
the space charge tune shift.
Positive and negative wakes are presented separately, no chromaticity.}
\end{figure}
%%%%%%%%%%%%%%%%%%%%%%%%%%%%%%%%%%%%%%%%%%%%%%%%%%%%%%%%%%%%%%%   Fig.7
%
%
%%%%%%%%%%%%%%%%%%%%%%%%%%%%%%%%%%%%%%%%%%%%%%%%%%%%%%%%%%%%%%%   Fig.8
\begin{figure}[h!]
\vspace{10mm}
\begin{center}
\includegraphics[width=85mm]{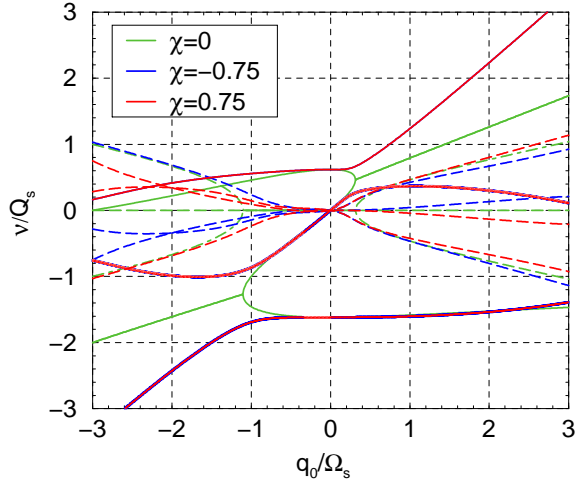}
\end{center}
\caption{Eigentunes of rectangular bunch with $\Delta Q_0=Q_s$ and 
chromaticity.  
Solid/dashed lines represent real/imaginary parts of the tune.
Some red and blue solid lines coalesce because real part of the tune is even 
function of chromaticity.}
\end{figure}
%%%%%%%%%%%%%%%%%%%%%%%%%%%%%%%%%%%%%%%%%%%%%%%%%%%%%%%%%%%%%%%   Fig.8
%

\newpage
\subsection{Discussion}

The most unstable TMCI mode with space charge was considered in Ref.~\cite{BL}
where expansion of eigenfunctions in terms of azimuthal and radial 
modes has been applied. 
Fig.~1 of the paper gives an example of negative wake which
is twice the size needed to produce instability without space charge:
$W/|W_{\rm thresh}(0)|=-2$.
The results depend on number of the basis modes which is characterized by the
number $m_{\rm max}$.
According them, the instability disappears at rather large 
$\Delta Q_{\rm sc}/Q_s$, and threshold values of this parameters are:
0.85 at $m_{\rm max}=1$ (3 multipoles), 
0.5 at $m_{\rm max}=5$ or 10 (21 multipoles and up to 5 radial modes in 
the last case).
My Eq.~(52) provide reasonably close parameters of the threshold:
$\Delta Q_0/Q_s=1$ at $q_0/Q_s=-1.14$ (point $X=1,\,Y=-1.14$ in Fig.~7).  

However, there is a profound disagreement at the larger space charge 
tune shift. 
A monotonous behavior of the threshold follows from my paper, and this
statement does not contradict Fig.~1 with $m_{\rm max}=1$ and 5.
However, new region of instability at $\Delta Q>2.2$ is predicted 
by the figure with $m_{\rm max}=10$.   
This problem is pointed but not explained in \cite{BL} (end of Sec.II).
  
Analytical and numerical investigation of the model with hollow bunch 
in square potential well is also performed in Ref.~\cite{BL}.
Being compared with Fig.~1, the results agree with option $m_{\rm max}=1$ 
better than with multi-modes approach.
At least, Fig.~13 obtained by numerical solution of differential equation 
demonstrates a monotonous behavior of the threshold, without additional 
instability regions at higher $\Delta Q_0$. 
However, analytical expression for $\Delta Q_x$ in page 10 gives a way of
head-tail instability even without chromaticity and space charge. 
I guess this statement is in a conflict with commonly accepted 
point of view \cite{CHAO, NG, HAND}.
My Eq.~(53) has another appearance and does not suffer from this shortcoming. 

My results correlate well with the multi-particle simulations presented 
in Ref.~\cite{BLA}.
For example, Fig.~1 of this paper resembles my Fig.~6 and allows to find 
TMCI thresholds of square bunch at $\Delta Q_0/Q_s=4$. 
According it, the relative wake strength is:
$W_{\rm thresh}/|W_0|\simeq -9$ for negative wake, and 0.2 -- 0.3 for positive one 
where $W_0$ is the threshold value for no space charge (it is difficult to get 
more exact numbers from the plot).
In terms of my paper $W_{\rm thresh}/|W_0|=(q_0)_{\rm thresh}/(0.57 Q_s)$ 
which value should be $-8.6$ or 0.2, according to my Fig.~7.
Another example is provided by Fig.~3 \cite{BLA} where threshold 
value of the wake is presented as a function of $\Delta Q/Q_s$.
The curves for square bunch coincide with my Fig.~7 not only in shape 
but also quantitatively.
Indeed, it follows from Fig.~3 that $W_{\rm thresh}/|W_0|\simeq-6$ 
at $\Delta Q_0/Q_s=3$. 
In terms of my paper, it means $q_{\rm thresh}/Q_s=-6\times0.57\simeq-3.4$ 
while the value -3.3 follows from Fig.7.
Positive wake thresholds are in a good consent as well being presented in 
these figures.
However, the agreement is not so close for smooth bunch.
According to Fig.~3 \cite{BLA}, thresholds of the smooth and square 
bunches have a similar behavior at $\Delta Q_{\rm sc}/Q_s<3$. 
The similarity is especially obvious if the averaged across the bunch value 
is used as the argument 
($\Delta Q_0=0.43\Delta Q_{\rm sc}$ for this distribution).  
However, the results come apart at larger tune shift  because non-monotonous 
behavior of the smooth bunch is shown in Ref.~\cite{BLA}.

%
%%%%%%%%%%%%%%%%%%%%%%%%%%%%%%%%%%%%%%%%%%%%%%%%%%%%%%%%%%%%%%%   SEC 8

\section{Conclusion}

%%%%%%%%%%%%%%%%%%%%%%%%%%%%%%%%%%%%%%%%%%%%%%%%%%%%%%%%%%%%%%%   SEC 8
%

The theory of a single bunch transverse instability is advanced in the paper 
by development of 3-modes model for the most unstable bunch modes.
The dispersion equation is presented in form of 3rd order algebraic 
equation which includes chromaticity and space charge, and can be used
with any bunch shape and wake field form.   

The known TMCI and head-tail instability appear in the theory as 
the limiting cases.
It is shown that a distinct boundary between them exists only at zero 
chromaticity representing the TMCI threshold in the case.
Generally, the TMCI appears more or less smoothly against the head-tail 
background without a pronounced demarcation line.
 
The results depend very slightly on the bunch shape so that  simple bunch 
models can be successfully used to analyze the stability limits. 
For example, difference of the TMCI thresholds is less of 1\% 
for so far models as hollow and Gaussian bunches, 
if they have the same  rms length and space charge is negligible.

In contrast with this, the tunes essentially depend on 
the wake form.
Several cases are investigated in the paper including arbitrary rectangular 
and resistive wall wakes. 
Comparison of the single bunch and multibunch/multiturn effects 
is realized in the last case.

Space charge tune shift is included in the consideration
at arbitrary relation of the shift to the synchrotron tune.
It is shown that the space charge effect depends on the wake sign: 
it increases the instability threshold if the wake is negative,
and decreases it at positive wakes.
Simple analytical formulae are presented for the instability threshold 
and growth rate.
They coincide well with the known expressions in the limiting cases 
though generally there are some divergences which are discussed 
in the paper.

%
%%%%%%%%%%%%%%%%%%%%%%%%%%%%%%%%%%%%%%%%%%%%%%%%%%%%%%%%%%%%%%%

%%%%%%%%%%%%%%%%%%%%%%%%%%%%%%%%%%%%%%%%%%%%%%%%%%%%%%%%%%%%%%%
%

\section{Appendix: Derivation of Eq.~(29)}

It follows from Eq.~(25) and (28) at $Y_0=1$, 
%
%%%%%%%%%%%%%%%%%%%%%%%%%%%%%%%%%%%%%%%%%%%%%%%%%%%%%%%%%%%%%%%   
$$
 \alpha^2 = 8\pi^3\int_0^\infty F(A)\,AdA\, 
 \biggl[\int_0^\infty  K_{0,1}(A',A)F(A')A'\,dA'\,\biggr]^2
$$
%%%%%%%%%%%%%%%%%%%%%%%%%%%%%%%%%%%%%%%%%%%%%%%%%%%%%%%%%%%%%%%   
%
The substitution of $K_{0,1}$ from Eq.~(21) results in
%
%%%%%%%%%%%%%%%%%%%%%%%%%%%%%%%%%%%%%%%%%%%%%%%%%%%%%%%%%%%%%%%   
\begin{eqnarray}
 \alpha^2 = \frac{128}{\pi}\int_0^\infty F(A)\,\frac{dA}{A}
 \biggl[\,\int_0^\infty F(A')A'dA' \int_{0}^{A'}
 \sqrt{\frac{A^2-\theta_{A}^2}{A'^2-\theta^2}}\,d\theta\,\biggr]^2\nonumber
\end{eqnarray}
%%%%%%%%%%%%%%%%%%%%%%%%%%%%%%%%%%%%%%%%%%%%%%%%%%%%%%%%%%%%%%%   
%
where $\,\theta_A={\rm min}\{\theta,A$\}.
Changing sequence of the integrals obtain
%
%%%%%%%%%%%%%%%%%%%%%%%%%%%%%%%%%%%%%%%%%%%%%%%%%%%%%%%%%%%%%%%   
\begin{eqnarray}
 \alpha^2 = \frac{128}{\pi}\int_0^\infty F(A)\,\frac{dA}{A}
 \biggl[\,\int_0^{A}\sqrt{A^2-\theta^2}\,d\theta  
 \int_\theta^\infty \frac{F(A')A'\,dA'}{\sqrt{A'^2-\theta^2}}\,\biggr]^2
 \nonumber
\end{eqnarray}
%%%%%%%%%%%%%%%%%%%%%%%%%%%%%%%%%%%%%%%%%%%%%%%%%%%%%%%%%%%%%%%   
%
The last integral is $\rho(\theta)/2$ so the expression is reducible 
in the form
%
%%%%%%%%%%%%%%%%%%%%%%%%%%%%%%%%%%%%%%%%%%%%%%%%%%%%%%%%%%%%%%%   
$$
 \alpha^2 = \frac{8}{\pi}\int_0^\infty F(A)A^3\,dA
 \biggl[\,\int_{-1}^1\rho(A\xi) \sqrt{1-\xi^2}\,d\xi\,\biggr]^2
$$
%%%%%%%%%%%%%%%%%%%%%%%%%%%%%%%%%%%%%%%%%%%%%%%%%%%%%%%%%%%%%%%   
%
which can be transformed in Eq.~(29) by the substitution $\xi=\cos\phi$.

\end{document}